\documentclass[11pt]{article}
\pdfoutput=1  
\usepackage{amsmath,amssymb}

\usepackage{epsfig}  
\usepackage{graphicx}               
\usepackage{url}
\usepackage{hyperref}

\usepackage{color}

\usepackage{epstopdf}

\newcommand{\nc}{\newcommand}  

\nc{\beq}{\begin{equation}}  
\nc{\eeq}{\end{equation}}  
\nc{\beqa}{\begin{eqnarray}}  
\nc{\eeqa}{\end{eqnarray}}  
\nc{\bea}{\begin{eqnarray}}  
\nc{\eea}{\end{eqnarray}}  
\nc{\ra}{\rightarrow}  
\nc{\Tr}{{\rm Tr}}
\nc{\slsh}{\slash\hspace*{-0.22cm}}

\def\be{\begin{equation}}
\def\ee{\end{equation}}
\def\bea{\begin{eqnarray}}
\def\eea{\end{eqnarray}}
\def\bit{\begin{itemize}}
\def\eit{\end{itemize}}

\newcommand{\gsim}{ \mathop{}_{\textstyle \sim}^{\textstyle >} }
\newcommand{\lsim}{ \mathop{}_{\textstyle \sim}^{\textstyle <} }

\newcommand{\gev}{{\rm\; GeV}}

\newcommand{\missET}{\slash{\hspace{-2.5mm}E}_T} 

\newcommand{\met}{E\!\!\!/_T}

\def\to{\rightarrow}

\title{  
\vspace*{-2.3cm}  
\begin{flushright}  
\normalsize{  
SLAC-PUB-14844\\
SCIPP-11/10\\
NSF-KITP-11-272
  }  
\end{flushright}  
\vspace{1.5cm}  
\Large  
\textbf{
Measuring the Invisible Higgs Width at the 7 and 8 TeV LHC
\\
}\vspace*{1.0cm}   
}

\author{Yang Bai$^{a}$, Patrick Draper$^{b}$ and Jessie Shelton$^{c}$
\vspace{5mm}
\\
$^{a}$ \normalsize\emph{SLAC National Accelerator Laboratory, 2575 Sand Hill Road, Menlo Park, CA 94025, USA} \\
$^{b}$  \normalsize\emph{Santa Cruz Inst. for Particle Physics, Univ. of California, Santa Cruz, CA 95064, USA} \\
$^{c}$ \normalsize\emph{Department of Physics, Sloane Laboratory, Yale University, New Haven, CT, 06520, USA}
}

\date{}

\begin{document}  
\setcounter{page}{0}  
\maketitle  

\vspace*{1cm}  
\begin{abstract} 
  The LHC is well on track toward the discovery or exclusion of a
  light Standard Model (SM)-like Higgs boson.  Such a Higgs has a very
  small SM width and can easily have large branching fractions to
  physics beyond the SM, making Higgs decays an excellent opportunity
  to observe new physics.  Decays into collider-invisible particles
  are particularly interesting as they are theoretically well
  motivated and relatively clean experimentally.  In this work we
  estimate the potential of the 7 and 8 TeV LHC to observe an
  invisible Higgs branching fraction.  We analyze three channels that
  can be used to directly study the invisible Higgs branching ratio at
  the 7 TeV LHC: an invisible Higgs produced in association with (i) a
  hard jet; (ii) a leptonic $Z$; and (iii) forward tagging jets. We
  find that the last channel, where the Higgs is produced via weak
  boson fusion, is the most sensitive, allowing branching fractions as
  small as 40\% to be probed at 20 inverse fb for masses in the range
  between 120 and 170 GeV, including in particular the interesting
  region around 125 GeV.  We provide an estimate of the 8 TeV LHC
  sensitivity to an invisibly-decaying Higgs produced via weak boson
  fusion and find that the reach is comparable to but not better than the reach at
  the 7 TeV LHC. We further estimate the discovery potential at the 8
  TeV LHC for cases where the Higgs has substantial branching
  fractions to both visible and invisible final states.
\end{abstract}  
  
\thispagestyle{empty}  
\newpage  
  
\setcounter{page}{1}  
    
\baselineskip18pt   

\vspace{-3cm}

\section{Introduction}
\label{sec:intro}

As the LHC approaches the first 10 fb$^{-1}$ of 7 and 8 TeV data, we are
entering a new era in Higgs physics. We will soon know whether or not
a Standard Model (SM)-like Higgs boson exists.  It becomes then an
urgent and fascinating question to verify that any observed particle
does---or does not---have the couplings and properties expected of the
SM Higgs boson.  Recent results presented by Atlas and CMS suggest a
light SM-like Higgs may soon emerge from the data in the mass range
below 130 GeV \cite{CMS5ifbHiggs,ATLAS5ifbHiggs}.  For such a Higgs
boson, the smallness of the $b$ quark Yukawa coupling means that its
total SM width is tiny: at 125 GeV, $\Gamma_h$ is only 4 MeV. Thus, a
light SM-like Higgs boson is especially sensitive to the existence of
physics beyond the Standard Model: even weak couplings to new light
degrees of freedom can have an $\mathcal{O}(1)$ effect on the Higgs'
branching
ratios~\cite{Shrock:1982kd,Ellis:1988er,Ellwanger:2001iw,Ellwanger:2004gz,O'Connell:2006wi,Keung:2011zc}.

One particularly interesting candidate for new degrees of freedom into
which the Higgs might decay is dark matter
(DM)~\cite{Burgess:2000yq,Pospelov:2011yp}. Dark matter can couple to
the SM through the Higgs portal~\cite{McDonald:1993ex,Patt:2006fw},
which provides one of two leading interactions for physics in other
sectors to couple to the SM. At colliders, Higgs-to-DM decays are
invisible apart from large missing energy. The Higgs portal coupling
also feeds into the Higgs-mediated DM-nuclear cross section,
correlating the results of direct DM detection experiments with
searches for invisible Higgs-to-DM decays at colliders. This
correlation is not rigid, in the sense that a nonobservation in direct
detection does not imply that the Higgs-to-DM branching ratio is
small~\cite{Pospelov:2011yp}, but the two handles are complementary
and can be used together to constrain models for DM (see, for
example,~\cite{Mambrini:2011ik,Raidal:2011xk,He:2011de,Weihs:2011wp}).

More generally, there are many compelling reasons to imagine that new
light degrees of freedom exist and couple preferentially to the Higgs
rather than to other parts of the SM.  Several questions of
naturalness in the Minimal Supersymmetric SM (MSSM) can be remedied by
extending the MSSM Higgs sector with a singlet
superfield~\cite{Ellwanger:2009dp}. Higgs decays into either the
singlet or the singlino can then lead to cascade or invisible decays
of the Higgs (for a review see \cite{Chang:2008cw}).  Other examples
which have been analyzed recently include decays of the Higgs through
pseudo-Goldstone bosons to four jets~\cite{Bellazzini:2011et},
multiple SUSY-breaking sectors where the Higgs decays to two
Goldstini~\cite{Bertolini:2011tw}, and a general analysis of Higgs
descendants in~\cite{Cheung:2011aa}.  In such scenarios, the Higgs is
often produced with cross-sections close to the SM values, while the
branching fractions to BSM particles can be large.

Searching for non-standard decay channels of a SM-like Higgs at the 7
TeV LHC therefore is a flexible, generic, and extremely well-motivated
probe of physics beyond the SM.  It is also critical as a cross check
of the SM Higgs production cross-sections: a universal discrepancy
between the measured and expected Higgs cross-sections could be due
either to suppression of the production cross-section from mixing with
another degree of freedom, or suppression of the visible branching
fractions by decay to invisible or
buried~\cite{Bellazzini:2009xt,Falkowski:2010hi} final states.  The
tiny Higgs width in the low-mass region is too small to measure
through its line shape, as might be imagined for a heavy Higgs
\cite{Low:2011kp}.  The scenarios can only be distinguished by
directly observing the nonstandard decay mode of the Higgs boson.

Existing searches for an invisibly-decaying Higgs at
LEP~\cite{:2001xz} have set an upper bound on its mass at 95\%
confidence level (C.L.) of 114.4 GeV, assuming SM production and 100\%
Higgs decay into invisibles.  At hadron colliders, an invisible Higgs
signature is more difficult than at lepton colliders, as the inability
to reconstruct events makes it challenging to separate the signal from
the the large physics backgrounds coming from $Z$ (and $W$), which
live at a similar mass scale as the signal.  However, the invisible
decay mode is still comparatively easy to detect. Large missing energy
is a relatively clean signal of electroweak physics, cleaner than, for
example, $b $-jets, and a signal in this channel may even be
accessible in the 7 TeV LHC run.  Both
theoretical~\cite{Frederiksen:1994me,
  Eboli:2000ze,Godbole:2003it,Davoudiasl:2004aj, Zhu:2005hv,
  Dedes:2008bf, Gopalakrishna:2009yz} and
experimental~\cite{ATLASHiggsInvisibleStudy,CMSHiggsInvisibleStudy,Cavalli:2002vs}
studies of an invisible Higgs at the 14 TeV LHC have been performed,
focusing on Higgs production from weak vector boson fusion (WBF) or in
association with the $Z$ boson. Higgs production from gluon fusion
plus one additional initial state radiation (ISR) jet has not been
considered a promising channel at the LHC.

Our aim in the current work is to update these studies and estimate
the sensitivity of the 7 TeV and 8 TeV LHC to an  invisibly decaying Higgs.  We
first assess existing bounds on an invisible Higgs.  The most
stringent constraints are indirect, coming from the reinterpretation
of SM Higgs exclusions as lower bounds on an invisible branching
fraction, assuming SM Higgs production.  There are also direct
constraints arising from existing searches in both
mono-jet$+\met$~\cite{CMSmonojet, Aad:2011xw, Atlasmonojet} and $h\to
ZZ\to \ell^++\ell^-+\met$~\cite{ATLASHZZ2,CMSHZZ}, which we
translate into bounds on the Higgs production times invisible
branching fraction,
\begin{align}
 \mathcal{B}_{\rm inv}\equiv \sigma\times BR_{\rm
  inv}/\sigma_{SM}.
\end{align}
These searches are sensitive to a light
invisible Higgs, produced through $gg\to h j$ and $p p \to Z h$,
respectively. We find that the mono-jet searches currently provide
a better limit than the $Z+h$ channel, although a targeted
optimization of the $Z+h$ channel might change this
picture.

Having established existing limits, we study the most sensitive
channel for a light invisible Higgs, namely weak vector boson fusion,
$qq\to hqq$.  We perform a detailed detector-level study of the WBF
channel to estimate the 7 TeV and 8 TeV LHC reach.  Incorporating improved
strategies for controlling systematic errors, we find that invisible
rates down to $\mathcal{B}_{\rm inv}\approx 0.4$ can be
excluded at 95\% CL with 20 fb$^{-1}$ at 7 TeV and a similar reach is found at 8 TeV.

The layout of this paper is as follows. In Section~\ref{sec:model}, we
briefly discuss simple reference models which can give the Higgs an
appreciable invisible branching fraction without altering electroweak
symmetry breaking.  The remainder of the paper is dedicated to
estimating the sensitivity of the 7 and 8 TeV LHC to an invisible Higgs
decay mode. In Sections~\ref{sec:monojet} and~\ref{sec:associated} we
adapt existing monojet$+\met$ and $Z+\met$ searches to establish
current limits on $\mathcal{B}_{\rm inv}$ from these channels.  In
Section~\ref{sec:WBF} we perform a detailed study of the weak boson
fusion channel at the 7 TeV LHC, which we find is the most sensitive channel.  In
Section~\ref{section:wbf8}, we
additionally estimate the sensitivity of the 8 TeV LHC in this
channel.  In Section~\ref{sec:combination} we compare the invisible
search reach to the visible search reach and discuss the ability of
the 8 TeV LHC to constrain the hypothesis of an invisibly decaying
Higgs through combinations of visible and invisible
channels. Section~\ref{sec:conclusions} contains our conclusions.

\section{Simple Models for Invisible Higgs Decays}
\label{sec:model}

\begin{figure}
\begin{center}
\hspace*{-0.75cm}
\includegraphics[width=0.48\textwidth]{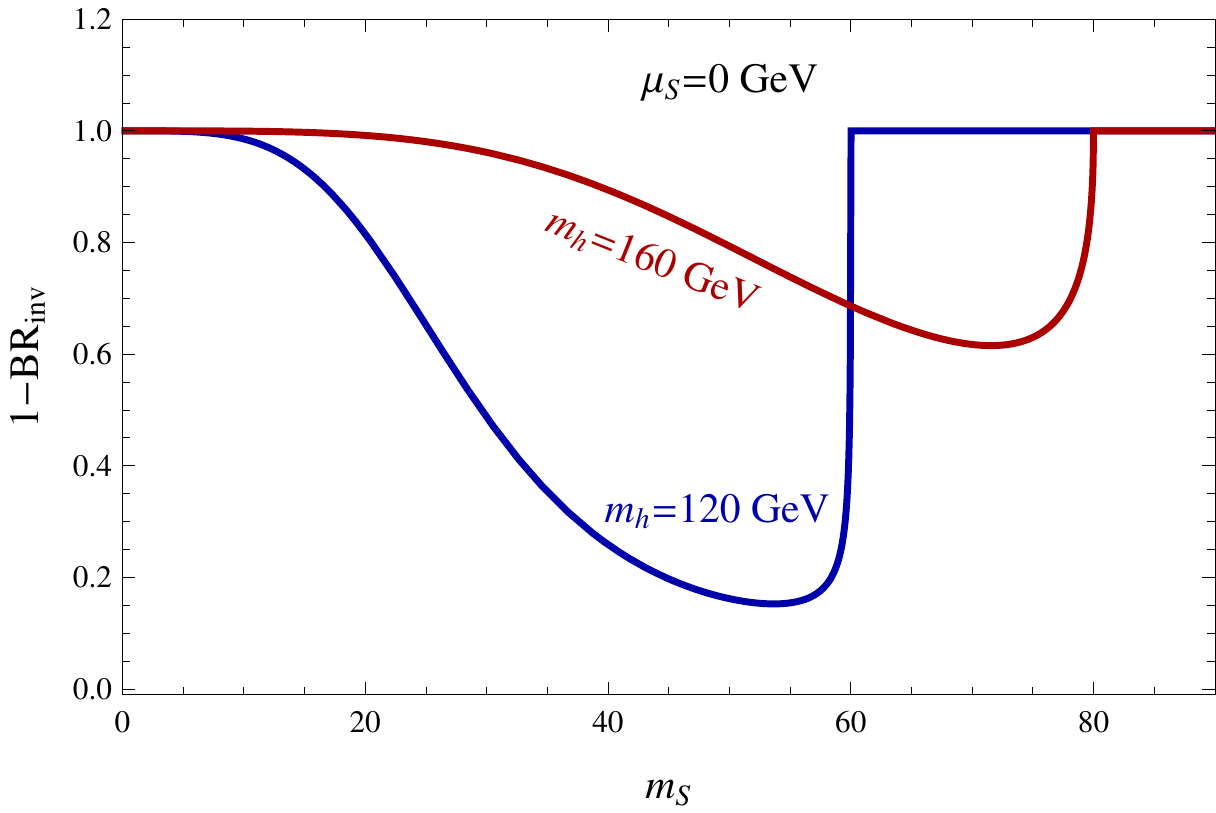} \hspace{3mm}
\includegraphics[width=0.48\textwidth]{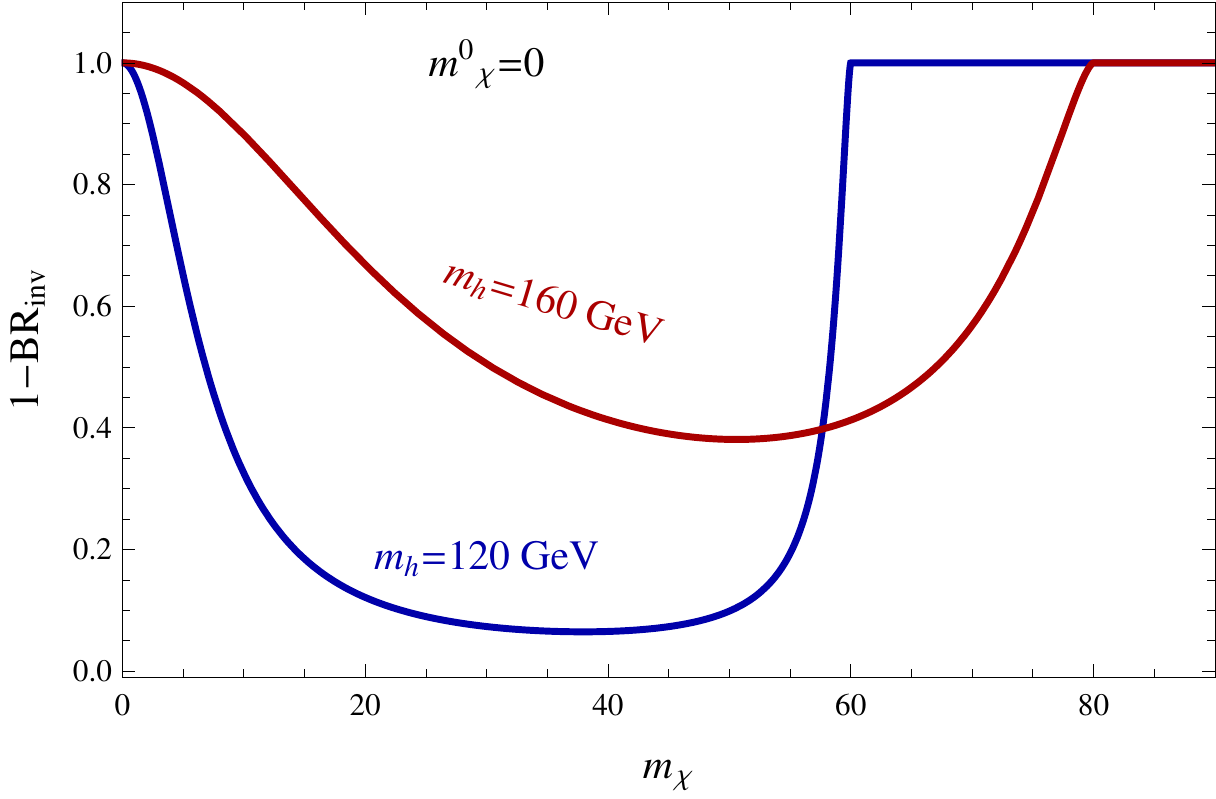} 
\caption{Left panel: the reduction of visible channel branching ratios after including a new invisible decay for the Higgs boson. The parameter $\mu_S$ is chosen to be zero, so the scalar $S$ has its mass proportional to the electroweak vacuum expectation value (VEV). Right: the same as the left panel but for a new fermion in the Higgs decay channel, with $m_\chi^0$ set to zero.}
\label{fig:ScalarBR}
\end{center}
\end{figure}

For simplicity, we consider a $\mathbb{Z}_2$ symmetry to protect one
new particle beyond the standard model from decaying. This new
particle could be a scalar $S$ (taken to be real) or a Dirac fermion
$\chi$, which are SM singlets. Here we neglect the case of a
new spin-one particle~\cite{Lebedev:2011iq}.
Starting from the scalar field~\cite{Burgess:2000yq, He:2011de}, we
have the following potential,
\beqa
V(H, S) =  - \,\mu^2\,HH^\dagger \,+\, \lambda \left( H H^\dagger\right)^2 \,+\, \frac{1}{2}\,\mu_S^2\,S^2 \,+\,\frac{1}{4} \,\lambda_S\,S^4 \,+\,  \kappa\,HH^\dagger\,S^2\,.
\label{eq:scalarpotential}
\eeqa
After electroweak symmetry breaking ($\langle H \rangle = v/\sqrt{2}$, $v=246$~GeV), the mass of the singlet becomes $m_S^2 = \mu_S^2 +
\kappa v^2$, which should be positive to prevent a vacuum
expectation value for $S$ and to keep the $\mathbb{Z}_2$ symmetry
intact.  The coupling modifying the Higgs decay is
\beqa
V(h, S) = \kappa\,v\, h \,S^2/2  \,+\, \cdots \,, 
\eeqa
which provides an invisible decay width for the Higgs boson,
\beqa
\Gamma(h \to S S) \,=\, \frac{\kappa^2\,v^2}{8\pi m_h}\,\sqrt{1- \frac{4 m_S^2}{m_h^2}} \,.
\eeqa
For a light Higgs boson, the dominant decay channel in the SM is into
two $b$ quarks. Since the bottom Yukawa coupling is small, it is
easy to modify the total width of Higgs boson and to suppress
the SM visible decay branching ratios. We show the
reduction of the visible branching ratios for two different Higgs
masses $m_h = 120, 160$~GeV in Fig.~\ref{fig:ScalarBR}. We
see from the left panel of Fig.~\ref{fig:ScalarBR} that the branching
ratios through visible channels can be
below 20\% of the branching ratios in the pure SM. In this plot, we
have chosen the bare scalar mass $\mu_S=0$ in
Eq.~(\ref{eq:scalarpotential}), so all of the $S$ mass comes from
electroweak symmetry breaking, and we use the
total SM widths calculated in
Ref.~\cite{Djouadi:2005gi}.

To generate Higgs decays to invisible fermions, we can extend the
renormalizable model to include couplings
\begin{align}
{\cal L} \supset \lambda_\chi S\overline{\chi}\chi - 2\mu S H^\dagger H + m^0_\chi \overline{\chi}\chi \,.
\end{align}
We include a bare mass for $\chi$ for completeness, but let us assume
for simplicity that $\lambda_S$ and $\kappa$ in
Eq.~(\ref{eq:scalarpotential}) (as well as any cubic coupling for $S$)
are small. Then EWSB generates a VEV for $S$, $\langle S\rangle
\approx \mu v^2/\mu_S^2$, and gives an additional mass contribution to
$\chi$,
\begin{align}
m_\chi^{E}=\frac{\lambda_\chi \mu v^2}{\mu_S^2} .
\end{align}
$S$ mixes with $h$, which allows $h$ to decay into
$\overline{\chi}\chi$. The width is given by \beqa \Gamma(h \to
\overline{\chi}\chi) \,=\, \frac{(m^E_\chi)^2\,m_h}{8\pi v^2}
\,\left(1-\frac{4 m_\chi^2}{m_h^2}\right)^{3/2} \,.
\label{eq:widthhchichi}
\eeqa
where $m_\chi$ is the total mass for $\chi$. We show the reduction of
visible channel branching ratios after including the
$\overline{\chi}\chi$ invisible decay for the Higgs boson in the right
panel of Fig.~\ref{fig:ScalarBR}.

\section{Direct Limits from Existing Searches}
\label{sec:direct}

In this section we consider limits on an invisible Higgs from existing
data, and briefly comment on future prospects in these channels. In
section~\ref{sec:monojet} we reinterpret searches in monojet$+\met$ to
establish sensitivity to an invisibly decaying Higgs produced through
gluon fusion.  In section~\ref{sec:associated} we reinterpret searches
for $h\to Z Z\to \ell^+\ell ^-\nu\bar\nu$ to obtain limits on an
invisible Higgs produced in association with a $Z$, $p p \to h Z$.  

\subsection{Monojet Searches}
\label{sec:monojet}

The simplest signal for invisible Higgs decays is monojet (or
monophoton) events, made by producing a Higgs in the gluon fusion
channel together with initial state radiation. Since the current
monojet searches provide a more stringent constraint on the invisible
Higgs rate $\mathcal{B}_{\rm inv}$ than the monophoton searches, we
will concentrate on the monojet signature and consider the existing
searches constraining $\mathcal{B}_{\rm inv}$.

Both CMS~\cite{CMSmonojet} and ATLAS~\cite{Aad:2011xw, Atlasmonojet}
have analyzed their monojet signatures at around 1 fb$^{-1}$ of
integrated luminosity. In this section, we take the results from
ATLAS~\cite{Atlasmonojet} to set a constraint on $\mathcal{B}_{\rm
inv}$. 
Ref.~\cite{Atlasmonojet} defines three sets of selection criteria
with different $p_T $ thresholds demanded for the leading jet
and the missing energy,
\begin {itemize}

\item {\texttt{LowPt}:} The leading jet is required to have $ p_{Tj_1}
  > 120$ GeV and $|\eta_{j_1}|<2$, together with $\missET > 120$ GeV.
  Other jets in the event ($|\eta_{j_2}|<4.5$) must be sufficiently
  soft, $p_{Tj_2} < 30$ GeV.

\item {\texttt{HighPt}:} The leading jet is required to have $
  p_{Tj_1} > 250$ GeV and $|\eta_{j_1}|<2$, together with $\met >
  220$ GeV.  Other jets in the event ($|\eta_{j_2}|<4.5$) must be
  softer than $p_{Tj_2} < 60$ GeV for the second-hardest jet, and
  $p_{Tj_3} < 30$ GeV for additional jets.  The missing energy must be
  separated from the second-hardest jet in azimuth, $\Delta\phi(j_2,
  \missET) < 0.5$.

\item  {\texttt{veryHighPt}:}  The leading jet is required to have $ p_{Tj_1} > 350$ GeV
   and $|\eta_{j_1}|<2$, together with $\missET > 300$ GeV. Other cuts on the subleading
  jets are the same as for the \texttt{HighPt} criteria.

\end {itemize}
 In all cases, leptons are rejected for electrons~(muons) above  $p_T > 20~(10)$ GeV
and  $|\eta| < 2.47~(2.4)$.  

\begin{figure}
\begin{center}
\hspace*{-0.75cm}
\includegraphics[width=0.48\textwidth]{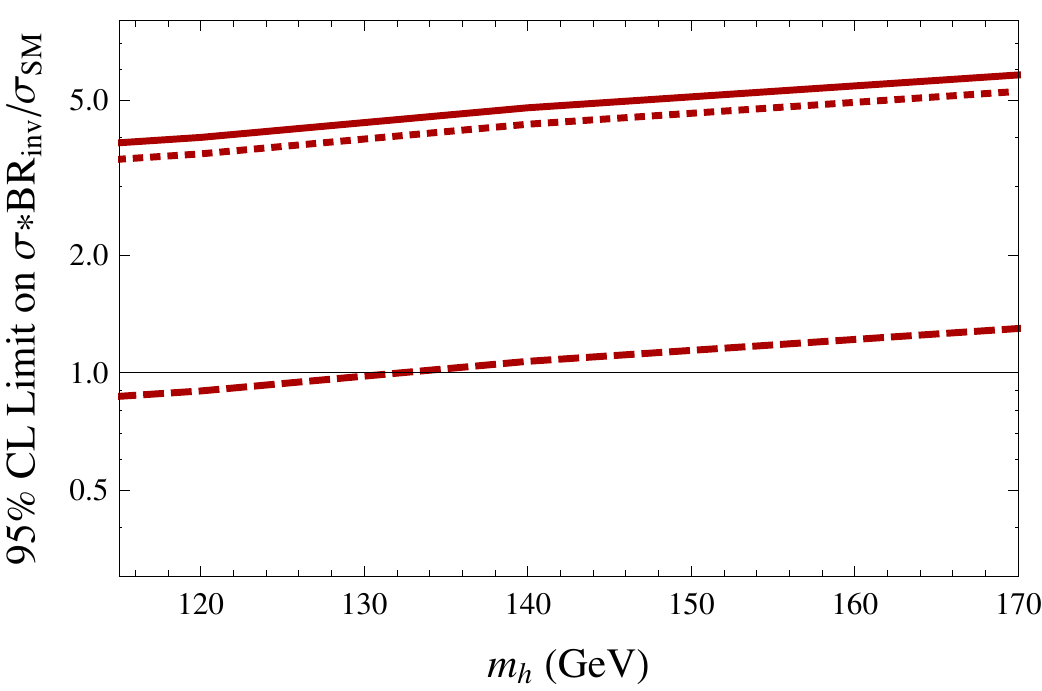} \hspace{3mm}
\includegraphics[width=0.48\textwidth]{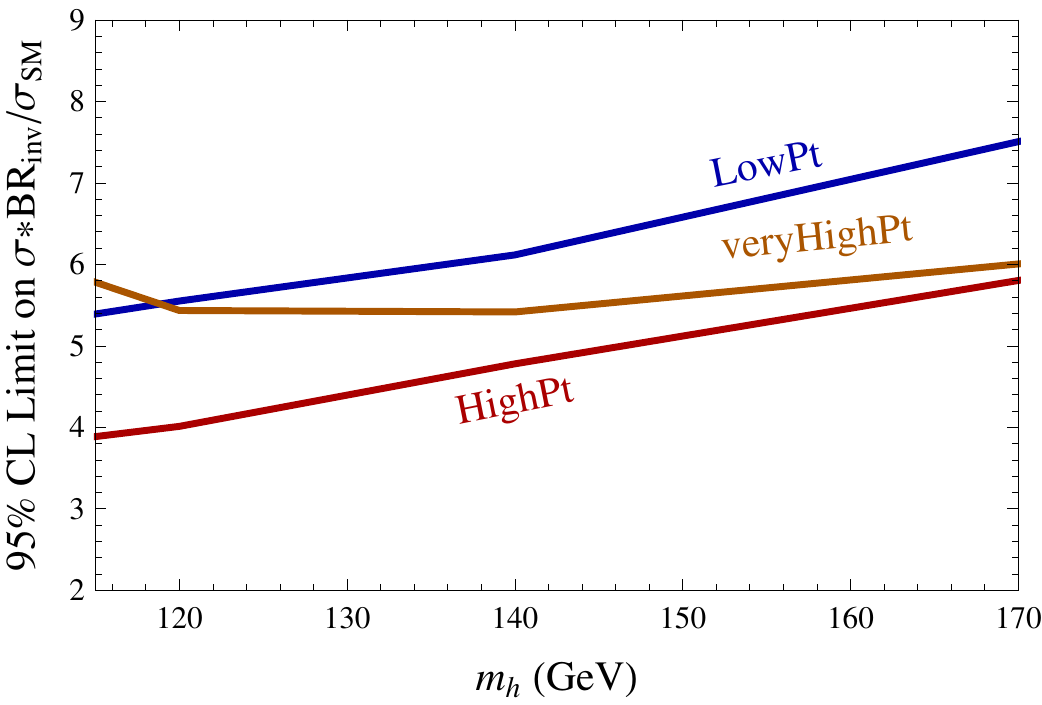} \hspace{3mm}
\caption{Left panel: the solid line is the current 95\% C.L. limit on
  the Higgs production times the invisible branching ratio from
  monojet searches at 1 fb$^{-1}$ with \texttt{HighPt} cuts. The
  dotted line is the projected limit at 20 fb$^{-1}$ assuming that
  the relative systematic error cannot be improved. The dashed line is
  the projected limit at 20 fb$^{-1}$ assuming that the relative
  systematic error can be improved and reduced alongside the
  statistical error by $1/\sqrt{\cal L}$. Right panel: A comparison of
  limits from three different cuts used in Ref.~\cite{Atlasmonojet}
  with 1 fb$^{-1}$ luminosity. }
\label{fig:monojet}
\end{center}
\end{figure}

We determine the efficiency of an invisibly decaying Higgs to pass
these cuts to obtain current and projected limits coming from this
channel.  Signal events are generated using
\texttt{FeynRules}~\cite{Christensen:2008py} in conjunction with
\texttt{MadGraph5}, and showered in
\texttt{PYTHIA}~\cite{Sjostrand:2006za}.  We use
\texttt{PGS}~\cite{PGS} to perform the fast detector simulation, after
modifying the code to implement the anti-$k_t$ jet-finding algorithm.
We generate our signal events $h$+jets using MadGraph's native MLM
matching~\cite{Mangano:2006rw} scheme and normalize to the inclusive
production cross section of $h$+jets at NNLO from
Ref.~\cite{Baglio:2010ae} to account for the large $K$-factor from
next-to-leading order (NLO) and next-to-next-to-leading order (NNLO)
corrections. In the present study, we neglect the contribution of signal events coming from weak boson fusion Higgs production, which contributes to the ATLAS monojet events at the 10\% level.

In Fig.~\ref{fig:monojet}, the solid red line shows the existing 95\%
confidence level (C.L.)  exclusion limit on $\mathcal{B}_{\rm inv}$
from the mono-jet searches at ATLAS with a 1 fb$^{-1}$
luminosity. Using the \texttt{HighPt} cuts, we add the statistical and
systematic errors in quadrature to set limits. To estimate the
potential improvement with luminosity, we show with the dotted line
the projected limits at 20 fb$^{-1}$ keeping the current relative
systematic error intact. The dashed line shows the projected limits
assuming that the relative systematic error can be reduced and scales
as $1/\sqrt{\mathcal{L}}$ as does the statistical uncertainty. In the
right panel of Fig.~\ref{fig:monojet}, we compare the sensitivities
of the three different sets of cuts. As one
can see from this panel, the \texttt{HighPt} cuts provide the best
exclusion limit~\footnote{Ref.~\cite{Englert:2011us} has recently used
  the {\texttt{LowPt}} cuts to set a limit. We emphasize that the
  {\texttt{HighPt}} analysis is more sensitive and provides a better
  starting point for an experimental search for invisible Higgs
  decays.}.  This can be understood as follows: for the \texttt{LowPt}
cuts the quoted systematic errors from multi-jets and non-collision
backgrounds are enormous and limiting; on the other hand, for a
sufficiently high $p_T$ cut, the major background $Z(\to
\nu\bar{\nu})$+jets has similar leading jet $p_T$ and $\missET$
distributions as the signal, so increasing the cuts decreases the
exclusion sensitivity.

Note that there is a large gap between the dotted line and the dashed
line in Fig.~\ref{fig:monojet}, indicating the dominance of systematic
errors in setting limits in this channel. The dashed line takes an
optimistic view that all relative systematic errors can be scaled with
luminosity as would be appropriate for uncertainties coming from
data-driven methods. This is not an unreasonable approximation as the
dominant uncertainties in Ref.~\cite{Atlasmonojet} are currently
limited by control region statistics, but of course not all systematic
errors will decrease with luminosity. 

We conclude by pointing out that the invisible Higgs could be further
separated from the dominant electroweak backgrounds by looking at the
composition of the associated jets.  Especially at lower $p_T$, the
leading diagrams contributing to the signal are dominated by gluon
radiation, while the $Z,W$ events are quark-enriched.  Unfortunately
the distinction is reduced at higher $p_T$, where the quark content is
increased for both signal and background.  However, since even at the
\texttt{LowPt} working point electroweak processes are by far the
dominant backgrounds, it is an interesting question whether additional
sensitivity can be achieved by lowering the $p_T$ cuts of the
\texttt{HighPt} and introducing quark-rejection variables
\cite{Gallicchio:2011xq}: the leading radiation in $h+j$ is $\sim 60
\%$ gluon-like at the \texttt{LowPt} working point, compared to $\sim
20\%$ gluon-like for $Z$ and $W$.  We consider this an interesting avenue
for future work.  However, we expect that this channel will have
difficulty achieving the sensitivity offered by $hZ$ and WBF.

\subsection{Associated Production of a Higgs with a $Z$}
\label{sec:associated}

We turn now to invisible Higgs decays in the so-called
``Higgsstrahlung'' channel, where the Higgs is produced in association
with a $Z$ boson. One promising channel is $q\bar{q} \to Z + h \to
\ell^+ \ell^- + \missET$.~\footnote{We have also checked the $W+h \to
  \ell +\missET$ channel and found that the signal acceptance
  efficiency is very small using the cuts in Ref.~\cite{Aad:2011yg},
  and that the final exclusion limit is much worse than that from the
  $q\bar{q} \to Z + h \to \ell^+ \ell^- + \missET$ channel; see also
  \cite{Gagnon:681580}.}  Although currently there is no invisible
Higgs search in this channel at the LHC, the standard Higgs search in
the channel $h \to Z Z \to \ell^+ \ell^- \nu \bar{\nu}$ provides an
identical final state and thus has the same backgrounds.

We consider the limits coming from the latest CMS search
\cite{CMSHZZ}, which uses an integrated luminosity of 4.6 fb$^{-1}$.
We summarize the set of cuts which offers the greatest sensitivity to
a light invisible Higgs,  namely the cuts used for $m_h=350$ GeV:
\begin {itemize}

\item  Events must contain two opposite sign, same flavor leptons with $p_T>20$ GeV which
 reconstruct a $Z$, $ |m_Z - m_{\ell \ell} | < 15$ GeV.

\item The reconstructed $Z$ must have $p_T>55$ GeV  and the missing energy must be greater than $\met > 95$ GeV.

\item  The transverse mass  of an assumed $ZZ$ final state,
$M_T^2=(\sqrt{p_{T,Z}^2+m_Z^2} + \sqrt{\met^2 + m_Z^2})^2 - (\vec{p}_{T,Z}+\vec{\met})^2$,
 should be in the range $298\gev< M_T < 393$ GeV.

\item Events with $b$ jets with $p_{T,b}>30$ GeV and $|\eta_b|<2.4$ or
  additional leptons with $p_T>10$ GeV are vetoed, as are events
  containing a jet with $p_T>30$ GeV which is too close in azimuth to
  the missing energy, $\Delta\phi(j, \missET) < 0.5$.  If the event
  contains no jets harder than 30 GeV, the azimuthal cut
  $\Delta\phi(j, \missET) < 0.5$ is applied to jets with $p_T> 15$
  GeV.

\end{itemize}
We  consider leptons within $|\eta|<2.5$.




%

\begin{figure}
\begin{center}
\hspace*{-0.75cm}
\includegraphics[width=0.48\textwidth]{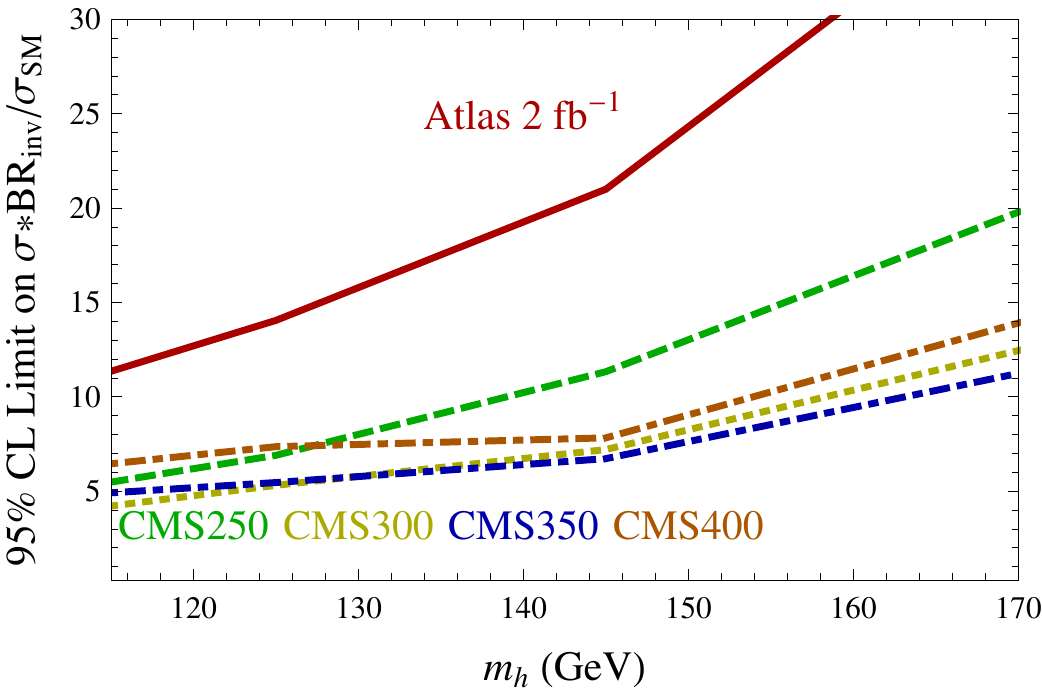}\hspace{3mm}
\includegraphics[width=0.48\textwidth]{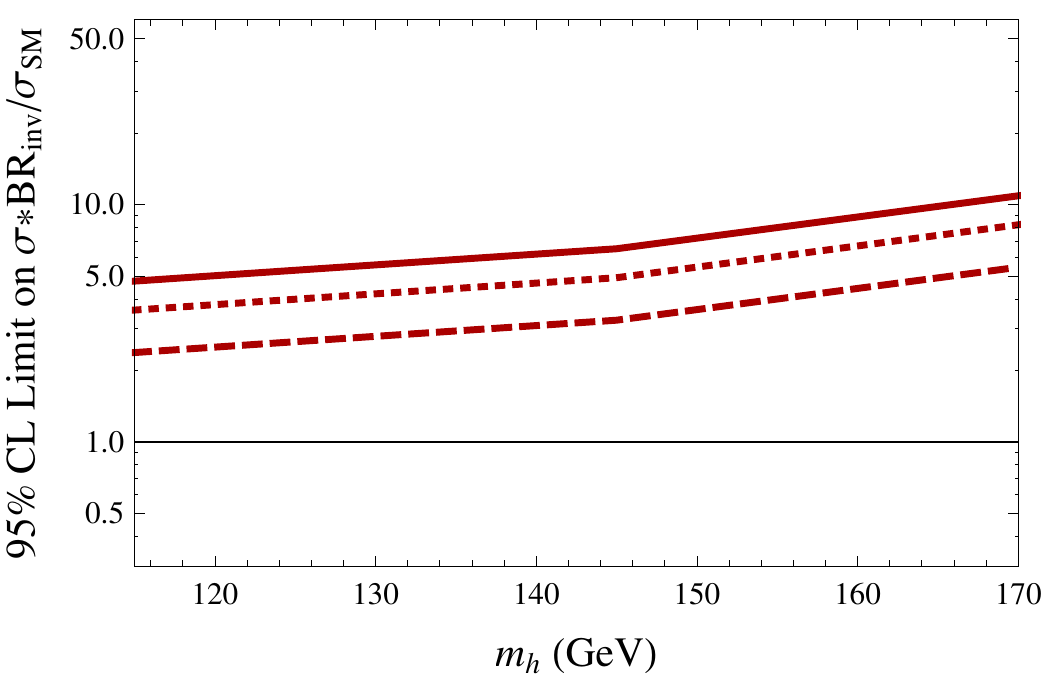}\hspace{3mm}
\caption{Left: Comparison of limits on Higgs production times
  invisible branching ratio coming from different sets of cuts
  employed in \cite{ATLASHZZ2,CMSHZZ}, interpreted as limits on $p p
  \to Zh\to \ell^+ \ell^- +$ invisibles.  Right: The most stringent of
  the current limits and future projections. The solid line is the
  current 95\% C.L. limit from the CMS search at 4.6 fb$^{-1}$. The
  dotted line is the projected limit at 20 fb$^{-1}$ by assuming that
  the relative systematic error cannot be improved. The dashed line is
  the projected limit at 20 fb$^{-1}$ assuming that the relative
  systematic error can be improved and reduced alongside the
  statistical error by $1/\sqrt{\cal L}$.  }
\label{fig:hz}
\end{center}
\end{figure}

We generate the signal events at tree-level using \texttt{MadGraph5}
and multiply the production cross section by a $K$-factor 1.3 from
NNLO results~\cite{Baglio:2010ae}. After applying the CMS cuts on our
signal events, we present the current constraint on $\mathcal{B}_{\rm
  inv}$ in the solid line of Fig.~\ref{fig:hz}. As one can see from
Fig.~\ref{fig:hz}, the invisible Higgs limit  coming from this search is
weaker than the limit coming from the monojet search.

%
These searches~\cite{ATLASHZZ2,CMSHZZ} are not targeted to the
invisible Higgs, and a dedicated analysis can improve the sensitivity;
see Ref.~\cite{Godbole:2003it,ATLASHiggsInvisibleStudy,Gagnon:681580}
for studies at 14 TeV.  However, we expect assisted production will
not be as sensitive to an invisible Higgs as weak vector boson fusion,
which we turn to next.


\section{LHC  Reach in Weak Boson Fusion at 7 TeV}
\label{sec:WBF}

The weak vector boson fusion (WBF) channel was previously shown to be
the most sensitive channel to an invisible Higgs at the 14 TeV LHC
\cite{Eboli:2000ze,ATLASHiggsInvisibleStudy,CMSHiggsInvisibleStudy}.
As the signal production proceeds mainly through the valence quark
PDFs while the largest backgrounds are also sensitive to $\bar q$ and
$g$ PDFs, it is unsurprising that the WBF channel remains the most
sensitive at 7 TeV.  Here we perform a study of the 7 TeV LHC's reach
in this channel, which should dominate the LHC's reach during the
low-energy run.  The aim of this section is to comprehensively update
the analysis of the WBF channel to 7 TeV, incorporating several
advances in signal and background computations, and to assess the
reach of the 7 TeV LHC for this signal.  All
results are detector-simulated using PGS.

The final state signature is characterized by two jets widely
separated in rapidity together with large missing energy.  The
required missing energy is large enough that we anticipate this
topology can be efficiently accepted through missing energy triggers.
The major backgrounds are $Z\to \nu\bar\nu$+jets from WBF; $W\to
\ell\nu$+jets from WBF, where the lepton is missed; $Z\to
\nu\bar\nu$+jets from QCD; $W\to \ell\nu$+jets from QCD with a missing
lepton; and finally mismeasured QCD.

To establish that we can neglect fake missing energy from QCD
multi-jet events, we estimate the scale of the mismeasured multi-jet
background by generating three-jet events using the Madgraph-Pythia
pipeline and running them through PGS.  Doing so, we find that the
mismeasured three-jet background can be suppressed by more than an
order of magnitude by demanding that the missing energy vector be
sufficiently distant in $\phi$ from the closest identified jet, 
\beq
\label{eq:metquality}
\mathrm{Min} (\Delta\phi_{j_i,\met}) > 0.5.  \eeq The multi-jet
contribution to the background falls off dramatically as a function of
missing energy, and is entirely subdominant above $\met \gsim 100$
GeV.\footnote{This accords with a similar conclusion obtained using
  Gaussian smearing in Ref.~\cite{Eboli:2000ze}; see also
  fast-detector simulated experimental studies in
  \cite{Cavalli:2002vs}.} While PGS only models the Gaussian portion
of the detector response, the non-Gaussian tails will act to enhance
configurations where a single mismeasured jet dominates the missing
energy, thereby increasing the probability that the resulting event
will fail the quality cut of Eq.~(\ref{eq:metquality}).  Since after
these two cuts the pure QCD background is orders of magnitude below
the other backgrounds, we subsequently neglect this background.

The large production cross-section of the Higgs through gluon fusion
means this contribution to the signal must be included, despite the
low efficiency of this process to pass the selection cuts.  We find
that this channel contributes approximately 10\% of the signal after
all selection cuts, broadly in accord with studies at 14 TeV
\cite{DelDuca:2001fn,Nikitenko:2007it,Cox:2010ug}.

The signal events as well as the physics backgrounds are generated in
\texttt{MadGraph5} and showered in \texttt{PYTHIA}.  Non-WBF processes
where the jets arise from QCD radiation ($gg\to h+$jets, $Z+$jets,
$W+$jets) are matched out to two jets using Madgraph's native MLM
matching scheme.  Detector simulation is done in \texttt{PGS} using
$R=0.5$ anti-$k_T$ jets.  After establishing our selection cuts, we
have checked that the {\it overall} efficiencies for leading order
$(W\to \tau\nu)+2j$ events are similar to those for fully matched
$(W\to \tau\nu)+$ jets events, and to increase our Monte Carlo
statistics we supplement our matched $(W\to \tau\nu)+$ jets
backgrounds with a sample of fixed $(W\to \tau\nu)+2j$ events.

Signal events produced through gluon fusion, $gg\to h$, are normalized
to the inclusive NLO production cross-sections tabulated in
\cite{Dittmaier:2011ti}.  Signal events produced through WBF are
multiplied by a constant $K$-factor of 0.95 (for all values of $m_h$)
which accounts for both QCD and electroweak corrections
\cite{Figy:2010ct}.  We use \texttt{VBFNLO} \cite{Arnold:2011wj} to
obtain a constant $K$-factor of 1.1 for the WBF $pp\to Zjj$, $pp\to
Wjj$ backgrounds.  The $W + $jets cross-section is normalized to the
inclusive cross-section of \cite{Berger:2010zx}.  Numerically this is
very close to using the LO cross-section for $W+2j$ from Madgraph and
applying the $K$-factor corresponding to the ratio of leading order
and next-to-leading order $W+2j$ cross-sections as found in
\cite{Berger:2010zx}.  Correspondingly, we normalize the the $Z+$jets
cross-section by multiplying the LO Madgraph cross-section with the
ratio of leading order and next-to-leading order $Z+2j$ cross-sections
found in \cite{Ita:2011wn}.

To estimate the $W$ backgrounds, we assume that events with
sufficiently hard and central leptons can be rejected. Specifically,
if an event contains a truth-level lepton with $|\eta_{\ell}|<2.5$, we
assume the event can be rejected if the lepton $p_T$ satisfies $p_T
>20$ GeV for electrons, $p_T>10$ GeV for muons, and $p_T>20$ GeV for
visible hadronic taus.  The resulting signal and background
cross-sections are displayed in Table~\ref{table:vbfXsec} for an
initial set of reference cuts, namely
\begin{equation}
\met > 90 \gev \phantom {spacer} \mbox{at least 2 jets with   } p_{Tj} >20\gev
       \phantom {spacer} \mathrm{Min}(\Delta\phi_{j_i,\met}) > 0.5
\end{equation}
in addition to the lepton veto.

Our main selection cuts are:
\begin{equation}
\met > 120 \gev \phantom {spac} \mbox{at least 2 jets with } p_{Tj} >30\gev,\phantom {spac} 
 M_{j_1j_2} > 1200 \gev,  \phantom {spac} \Delta\eta_{j_1j_2}>4.5.
\end{equation}
We also cut on the azimuthal angle of the tagging jets, requiring that
they not be back-to-back:
\begin{equation}
\Delta\phi_{j_1j_2}<1.5\,.
\end{equation}
The azimuthal angle is the one distribution which has a markedly
different shape for the Higgs than for other WBF processes
\cite{Eboli:2000ze}.  As this distribution probes the spin
correlations of the jets, it is not modeled faithfully by jets arising
from parton showers in Pythia. To model the efficiency for
background $W,Z+$ jets processes to pass the $\Delta\phi$ cut, we
consider fixed order $W,Z+2$ jet events generated in MadGraph, which
does retain spin information.  We find that the difference in
efficiency between fixed order and matched events decreases as
$m_{jj}$ increases, and is order 10\% in the regions of interest here.
Accordingly, we account for the lack of spin correlations in the
matched samples by multiplying the efficiencies for $W,Z+$ jets to
pass the $\Delta\phi$ cut by a factor of 1.1.

Lastly, we impose a central jet veto.  We show results for two different
implementations of the veto, 
\begin {enumerate}
\item requiring {\it any} central  jet ($|\eta| <2.5$),
 including the two tagging jets, to have  $p_T<40$ GeV, and 
\item
 requiring any central jet in addition to the two leading jets to have
$p_T < 25$ GeV.
\end{enumerate}   
The second and more traditional implementation of the jet veto offers
a better limit with only statistical uncertainties taken into account,
but the first implementation of the veto gives the best results after
systematic uncertainties are included according to our prescription.
Signal and background cross-sections through these cuts are displayed
in Table~\ref{table:vbfXsec}.

\begin{table}
\begin{center}
\begin{tabular}{|c|c|c|c|c|c|c|c|c|}
\hline
\hline
Cross section (fb) & $qqh$ & $hjj$ & $qqZ$  & $Zjj$ & $qqW $ & $Wjj $  \\
\hline 
\hline

 Reference cuts & 315 & 647 & 404 & 32600 & 465 & 32300 \\ 
\hline
 WBF  selection cuts & 14.1 & 1.86 & 6.82 & 24.8 & 7.30 & 17.9\\

 $\Delta\phi$ cut & 9.26 &  1.39 & 2.05 & 12.5 & 2.53 & 9.73\\

 jet veto (1) & 4.23 & 0.41 & 0.77 & 3.37 & 1.11 & 2.89 \\
 jet veto (2) & 7.25 & 0.67 & 1.23 & 7.58 & 1.81 & 6.08 \\

\hline
\hline
\end{tabular}
\end{center}
\caption{Cross-sections in fb for signal and background in the vector boson fusion Higgs search channel.
 Signal cross-sections are shown for $m_h=120$ GeV.   The cuts are described in the text.
 \label{table:vbfXsec}
}
\end{table}

Note that $W +$jets with a missing lepton, and in particular with a
missing $\tau$, is numerically as large as the background from the $Z$.
In particular $W\to\tau\nu +$jets contributes more than half of the
$W$ background.  Vetoing more aggressively on charged leptons is thus
one obvious avenue to maximize the sensitivity.  Another possible
avenue of improvement is lowering the missing energy cut, which at 120
GeV is slightly too hard for optimal significance.  We keep the
missing energy cut hard to be conservative about trigger acceptance,
forward jet resolution, and multi-jet background, but lowering this
threshold could well prove useful in this channel.

\begin{figure}
\begin{center}
\includegraphics[width=0.48\textwidth]{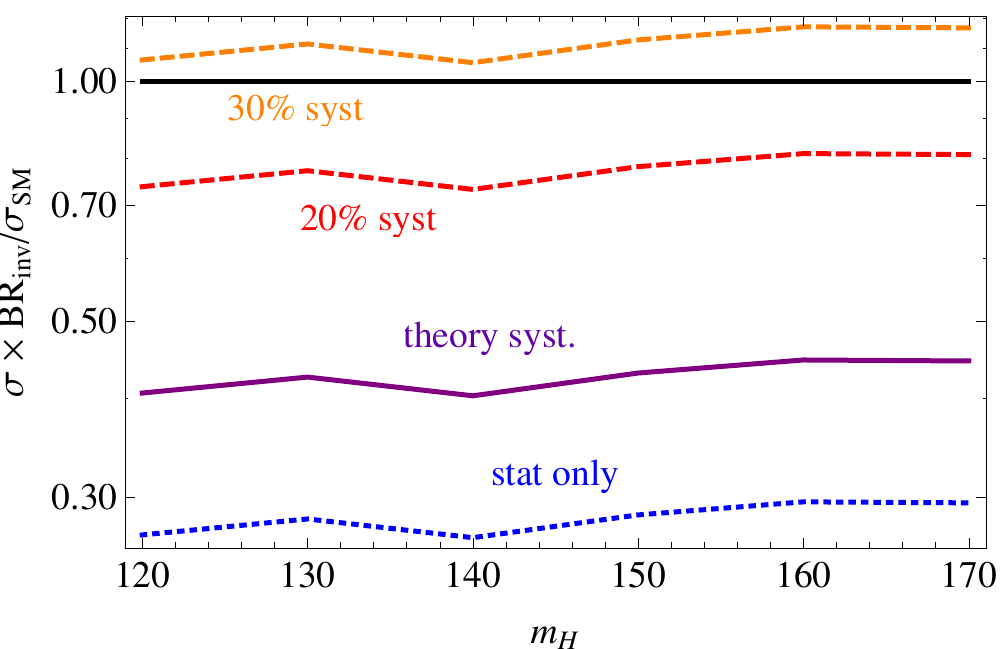} \hspace{3mm}
\includegraphics[width=0.48\textwidth]{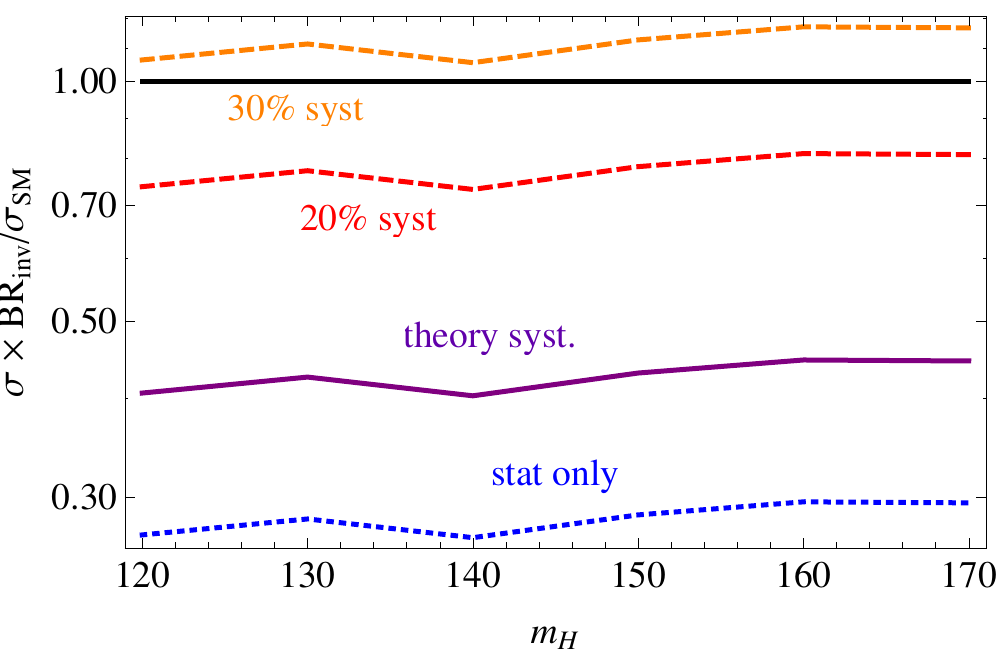} \\
\includegraphics[width=0.48\textwidth]{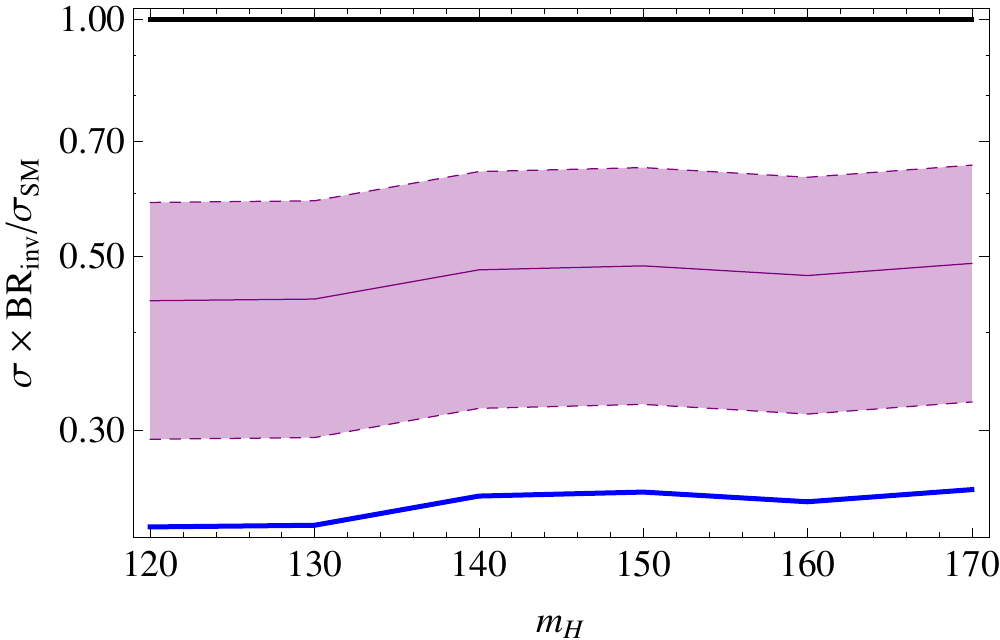} \hspace{3mm}
\includegraphics[width=0.48\textwidth]{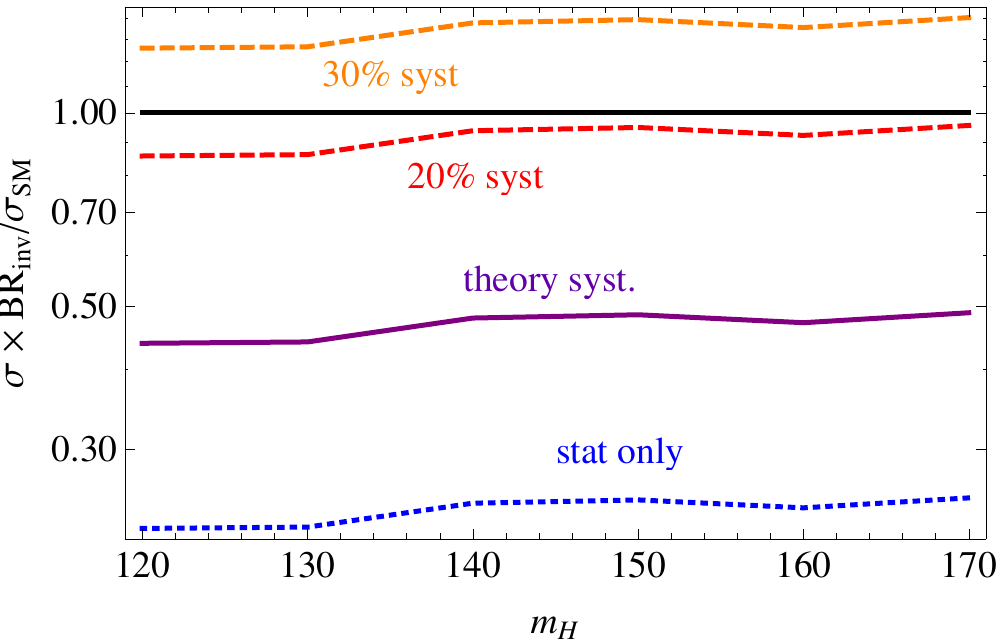} 
\caption{Left: Estimated 95\% CL limit on the invisible branching fraction of the
  Higgs with 20 fb$ ^ {-1}$ of data at the 7 TeV LHC.  The straight line at 1
  denotes a Higgs produced with SM cross-sections decaying entirely
  into invisibles.   Shaded bands show $\pm 1 \sigma$ with systematic 
  uncertainties assigned as described in the text. Right: Dependence of the 
  95\% CL limits on systematic uncertainties.  
  The top row shows results for jet veto (1) as described in the text; the bottom 
  row shows results for jet veto (2).
 \label{fig:vbflimit}}
\end{center}
\end{figure}

We plot the estimated 95\% CL exclusion reach of the 7 TeV LHC in
the left panel of
Figure~\ref{fig:vbflimit}.  Since no Higgs mass peak can be
reconstructed, this signal is purely a counting experiment, and
systematic uncertainties on the background cannot be neglected when
estimating the reach in this channel.  To reduce theoretical
uncertainties associated with modeling the $Z, W + $jets backgrounds,
the backgrounds can be modeled from control regions in the data.
Inverting the lepton selection criteria for $W + $jets furnishes a
large control sample. For $Z + $jets, the control sample provided by
$Z\to\ell ^ +\ell^-$ is statistically limited by the small leptonic
branching ratio of the $Z $.  A more promising technique is to
reweight $\gamma +$ jets, exploiting the stability of the ratio
$\sigma(\gamma +X)/\sigma (Z+ X)$
\cite{CMS-PAS-SUS-08-002,CMS-PAS-SUS-10-005,Bern:2011pa}.  We expect
that we are in the regime where this technique can be applied.  The
cut of Eq.~(\ref{eq:metquality}) requiring the missing energy to be
azimuthally separated from the jets ensures that we do not probe the
region of phase space where the photon amplitudes are dominated by the
collinear singularity, and the missing energy cut we apply is not
substantially different from those studied in \cite{Bern:2011pa}.
Using photons allows the $Z\to \nu\bar\nu$ contribution to the
background to be determined to below $ 10\% $.  We assume a $10\% $
theoretical uncertainty on $W,Z +$ jets when calculating our
systematic error bars.  The WBF production cross-sections for $W, Z $
are relatively stable \cite{Oleari:2003tc} and we assume a
$5\% $ theoretical uncertainty for these backgrounds.

The cuts employed in this analysis do not exploit all of the kinematic
information in the final state: the azimuthal distance between the
missing energy and the jets contains some distinguishing power
between signal and background. We have checked that cutting on
appropriate linear combinations of $\Delta\phi_{\met j_1}$ and
$\Delta\phi_{j_1j_2}$ offers a marginal improvement in the
reach. However, the azimuthal angle of the missing energy is subject
to larger experimental systematic uncertainties than the
(well-measured) azimuthal angles of the jets, so it is unclear whether
the improvement will persist after all experimental systematic
uncertainties have been taken into account.  We comment that since the
signal process in this channel is relatively well described by a LO
matrix element, a matrix element technique
\cite{Kondo:1988yd,Abazov:2004cs} seems promising as a way to fully
exploit the event kinematics and offer improved separation between
signal and background.  A fully accurate estimate of the 7 TeV LHC's
reach in this channel however requires a more accurate detector
simulation. In particular, the final state signature is subject to
uncertainties in the forward jet energy scale which are exacerbated by
pileup. Large experimental systematics on jet resolution could spoil
the sensitivity of a matrix element-based kinematic fitter.  The right
panel of Figure~\ref{fig:vbflimit} shows how the estimated limit
weakens as larger systematic uncertainties are included.  We consider
the results in this section strong motivation for a more detailed
investigation of this channel.

\subsection {LHC reach in weak boson fusion at 8 TeV}
\label{section:wbf8}

NLO calculations are not available at 8 TeV for all of the required
signal and background processes. Nonetheless, we can estimate the
reach of the 8 TeV LHC by taking the $K$-factors to be approximately
equal at 7 TeV and at 8 TeV. We generate signal and background WBF
processes at LO and normalize with the same constant $K$-factor used
at 7 TeV. We generate matched $W,Z,h+$ jets samples at 8 TeV, and
normalize to the 7 TeV NLO cross-sections, then rescale by the ratio
$\Delta = \sigma_{inc,8}^{(0)}/ \sigma_{inc,7}^{(0)}$.

The resulting signal and background
cross-sections are displayed in table~\ref{table:vbfXsec} for an
initial set of reference cuts, namely
\begin{equation}
\met > 90 \gev\,, \phantom {spacer} \mbox{at least 2 jets with } p_{Tj} >20\gev \,,
       \phantom {spacer} \mathrm{Min}(\Delta\phi_{j_i,\met}) > 0.5\,,
\end{equation}
in addition to the lepton veto.

At 8 TeV, the selection cuts can be tightened:
\begin{equation}
\met > 120 \gev \phantom {spac} \mbox{at least 2 jets with } p_{Tj} >35\gev,\phantom {spac} 
 M_{j_1j_2} > 1600 \gev,  \phantom {spac} \Delta\eta_{j_1j_2}>4.5,
\end{equation}
 in addition to the cut on the azimuthal angle of the tagging jets,
\begin{equation}
\Delta\phi_{j_1j_2}<1.5\,.
\end{equation}
We again apply a correction factor of 1.1 to the $W,Z+$ jets
backgrounds after the $\Delta\phi$ cut to better model the angular
distributions.  Finally, we apply a central jet veto, requiring no
additional central jets with $p_T < 20$ GeV.
Signal and background cross-sections through these cuts
are displayed in Table~\ref{table:vbfXsecat8}.

\begin{table}
\begin{center}
\begin{tabular}{|c|c|c|c|c|c|c|c|c|}
\hline
\hline
Cross section (fb) & $qqh$ & $hjj$ & $qqZ$  & $Zjj$ & $qqW $ & $Wjj $  \\
\hline 
\hline
 Reference cuts & 404  & 951 & 517 & 39100  & 594  & 49400  \\ 
\hline
 WBF  selection cuts &14.1  &  1.86 &  6.82 &  24.8 &  7.30 & 17.9  \\ 
 $\Delta\phi$ cut &  5.72 &  0.61 & 1.65  & 3.73  &  1.68 &  6.89 \\
 jet veto & 4.11 &0.32 & 1.27 & 2.40 & 1.30 & 4.20 \\
\hline
\hline
\end{tabular}
\end{center}
\caption{8 TeV cross-sections in fb for signal and background in the weak boson fusion Higgs 
search channel. Signal cross-sections are shown for $m_h=120$ GeV.   The cuts are described 
in the text.
 \label{table:vbfXsecat8}
}
\end{table}

 \begin{figure}
\begin{center}
\hspace*{-0.75cm}
\includegraphics[width=0.48\textwidth]{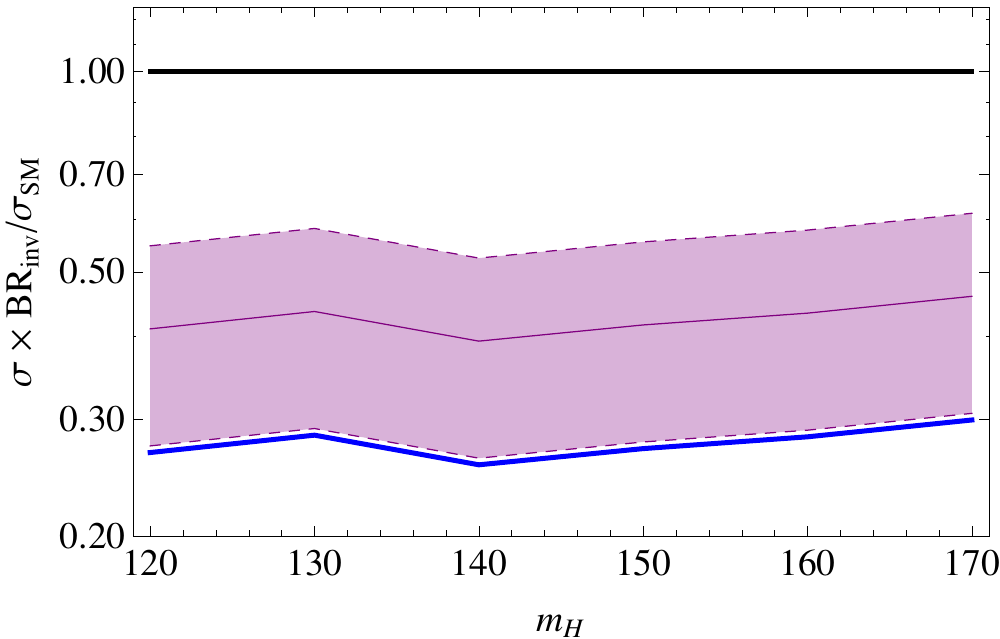} \hspace{3mm}
\includegraphics[width=0.48\textwidth]{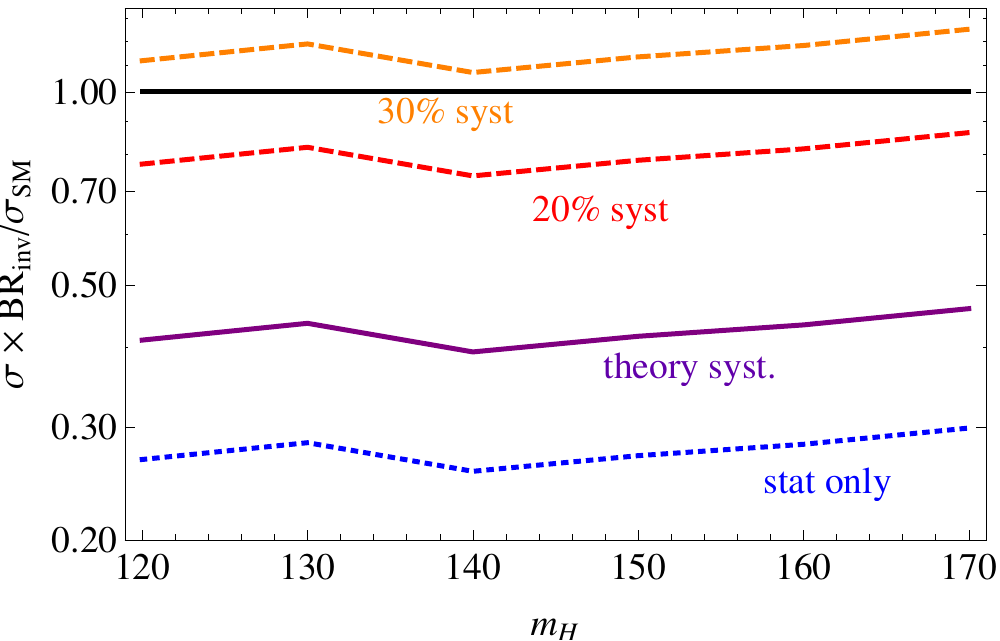} \hspace{3mm}
\caption{Left: Estimated 95\% CL limit on the invisible branching fraction of the
  Higgs with 20 fb$ ^ {-1}$ of data at the 8 TeV LHC.  The straight line at 1
  denotes a Higgs produced with SM cross-sections decaying entirely
  into invisibles.   Shaded bands show $\pm 1 \sigma$ with systematic uncertainties assigned as described in the text. Right: Dependence of the 95\% CL limits on systematic uncertainties.
  \label{fig:vbflimitat8}}
\end{center}
\end{figure}

Prospects at the 8 TeV LHC do not improve relative to the 7 TeV LHC
for the same amount of integrated luminosity.  This is unsurprising:
the signal, which arises dominantly from valence quark PDFs, does not
increase as quickly with $\sqrt{s}$ as do the background $W$ and $Z+$
jets processes.  Tightening the selection cuts to improve $S/\sqrt{B}$
further suppresses the contribution of $gg\to hjj$ to the signal
reach.  Thus in this channel the increased center of mass energy gives
little advantage.

\section{Combination with SM Higgs search channels}
\label{sec:combination}
\begin{figure*}[!t]
\begin{center}
\includegraphics[width=0.45\textwidth]{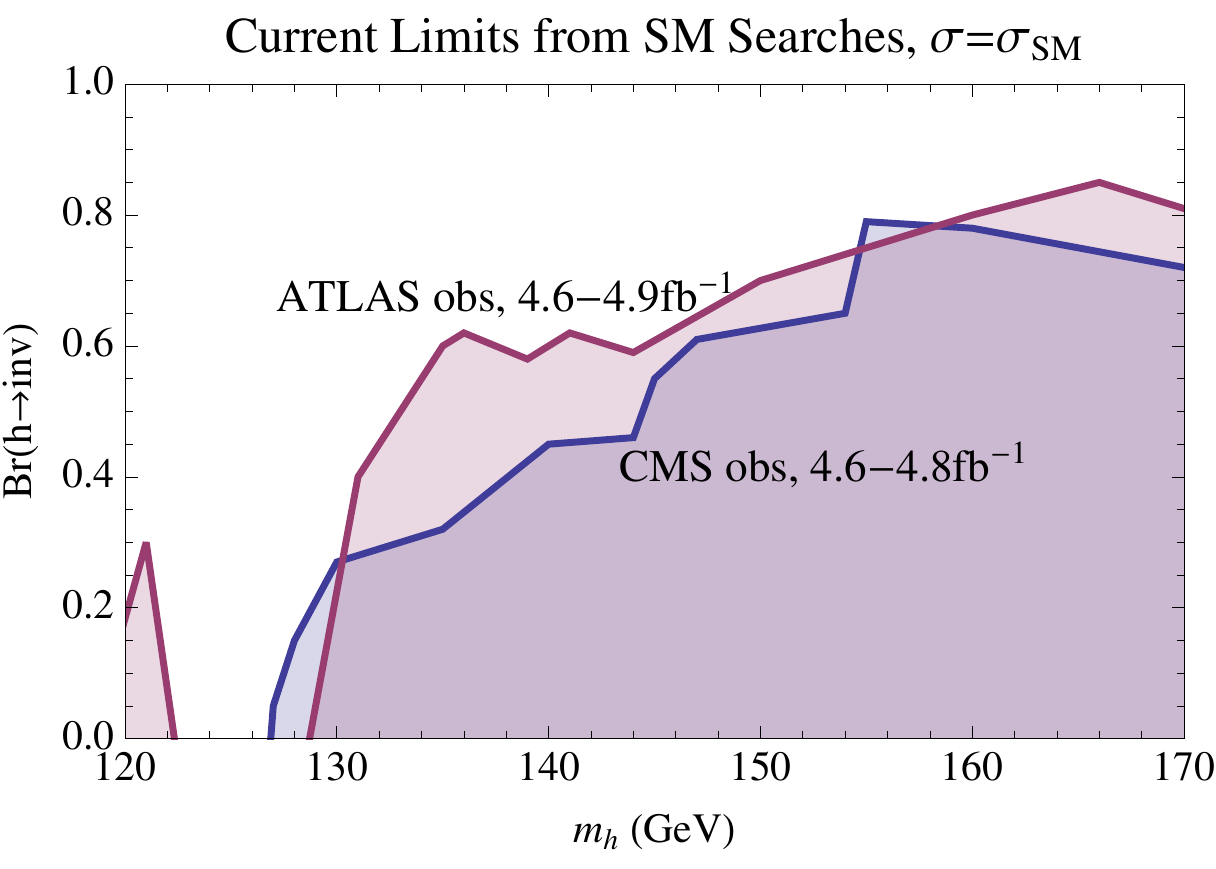}
\includegraphics[width=0.45\textwidth]{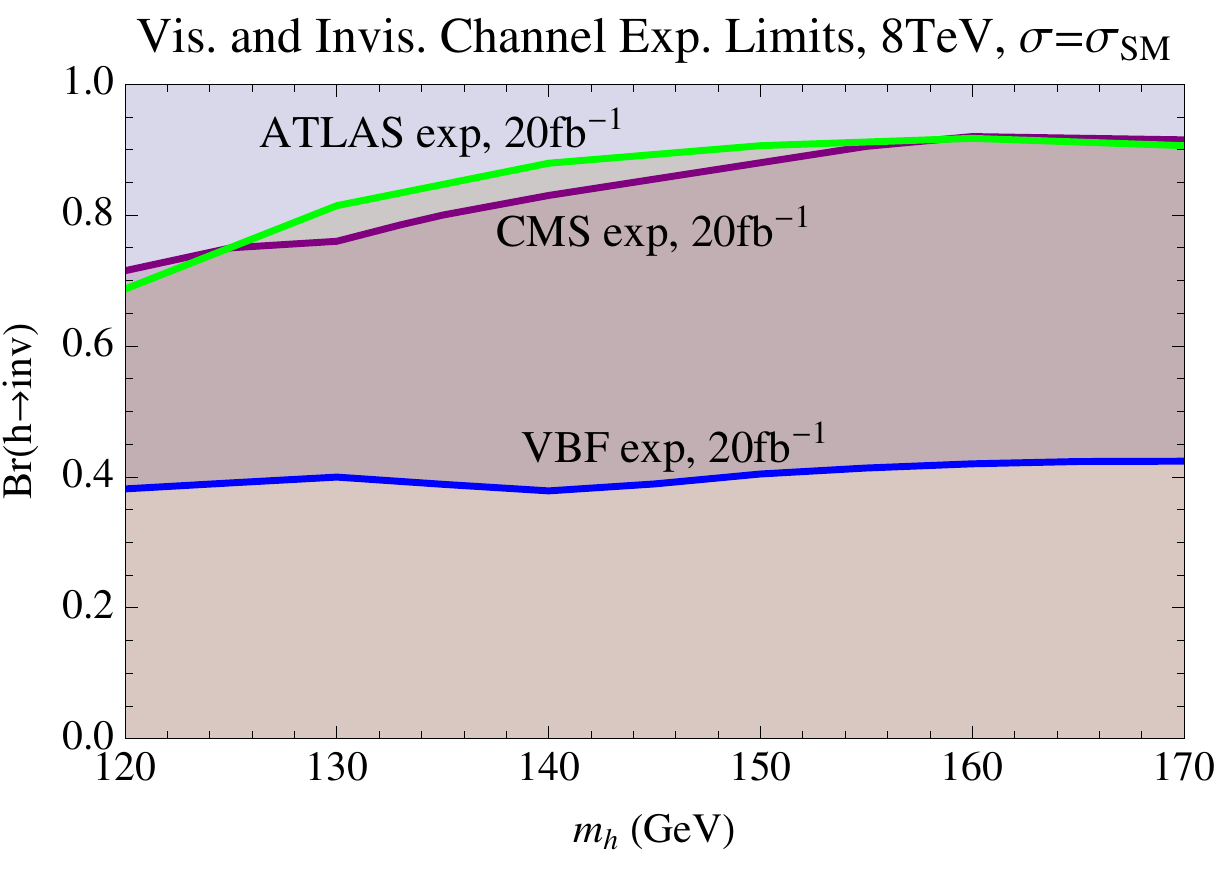}
\caption{Left: Present observed lower bound on the invisible branching
  ratio assuming the IHH. Right: Expected lower limit from visible
  channels and upper limit from the WBF invisible channel with
  20~fb$^{-1}$/channel at 8 TeV.}
\label{fig:currentlimit}
\end{center}
\end{figure*}

If no signal is observed in SM Higgs channels, it will be of interest
to analyze the channels probing invisible Higgs decays. We define the
invisible Higgs hypothesis (IHH) to be the case $\{\sigma=\sigma_{SM}$,
$BR(h\rightarrow \mbox{inv})+BR(h\rightarrow \mbox{vis})=1\}$, or
$\mathcal{B}_{inv}=BR(h\to \mbox{inv})$. Current SM Higgs limits can
then be reinterpreted as lower limits on $BR(h\rightarrow
\mbox{inv})$ in the IHH, as shown in the left-hand panel of
Fig.~\ref{fig:currentlimit}.  

If no signal is observed in the
invisible channels, then upper limits may be placed on the IHH,
providing complementary coverage as shown in the right-hand panel of
Fig.~\ref{fig:currentlimit} for 20 fb$^{-1}$/channel at 8 TeV (projections for visible channels are made with $\sqrt{\cal L}$ rescaling of the expected limits given in~\cite{ATL8,CMS8}.) 
For sufficiently large invisible widths, the total Higgs width can
begin to exceed the experimental resolution of the Higgs bump, which
would weaken the limits from the visible channels. This effect is
not large, and negligible in the very light ($m_h\lsim 130$ GeV) range of most
interest.  We use the expected limits from the visible channel searches
directly without modification.

Even better limits may be derived by combining the visible and
invisible SM Higgs searches. Allowing the cross section to float, one
can make the usual ``Brazil-band'' plots, where now the $y$-axis
parametrizes a space of models that differ from the IHH by a universal
rescaling $R$ of each channel's rate. Limits below $R=1$ exclude the
IHH for a given pair $\{m_h, BR(h\rightarrow \mbox{inv})\}$. Strong
limits will motivate a variety of interpretations not covered by the
IHH -- perhaps there is no Higgs, or the Higgs is buried, or the
production cross section is suppressed. Values of $R<1$ can be
interpreted as a minimum required universal suppression factor of the
cross section, or one minus the minimum required branching ratio into
``buried'' final states.

We can make estimates for the combined reach of SM+invisible search
channels with a quadrature combination of the statistical
significances, or inverse quadrature combination of the individual
limits on $R$,
\begin{align}
\left(\frac{1}{R^{95}}\right)^2\approx\left(\frac{1}{R^{95}_{\rm{vis}}}\right)^2+\left(\frac{1}{R^{95}_{\rm{inv}}}\right)^2\; .
\end{align}
Here $R^{95}$ denotes the 95\% CL limit on $R$.  Since the error bars
in each channel include systematic errors that are in general
correlated across channels, our combinations are approximate and
intended only to suggest the potential interplay between visible and
invisible channels. To estimate the search sensitivity to the visible Higgs
decay modes, we use the recent expected limits from CMS
\cite{CMS5ifbHiggs} as they combine several channels with (almost) uniform luminosity
 per channel.

\begin{figure*}[t!]
\begin{center}
\includegraphics[width=0.45\textwidth]{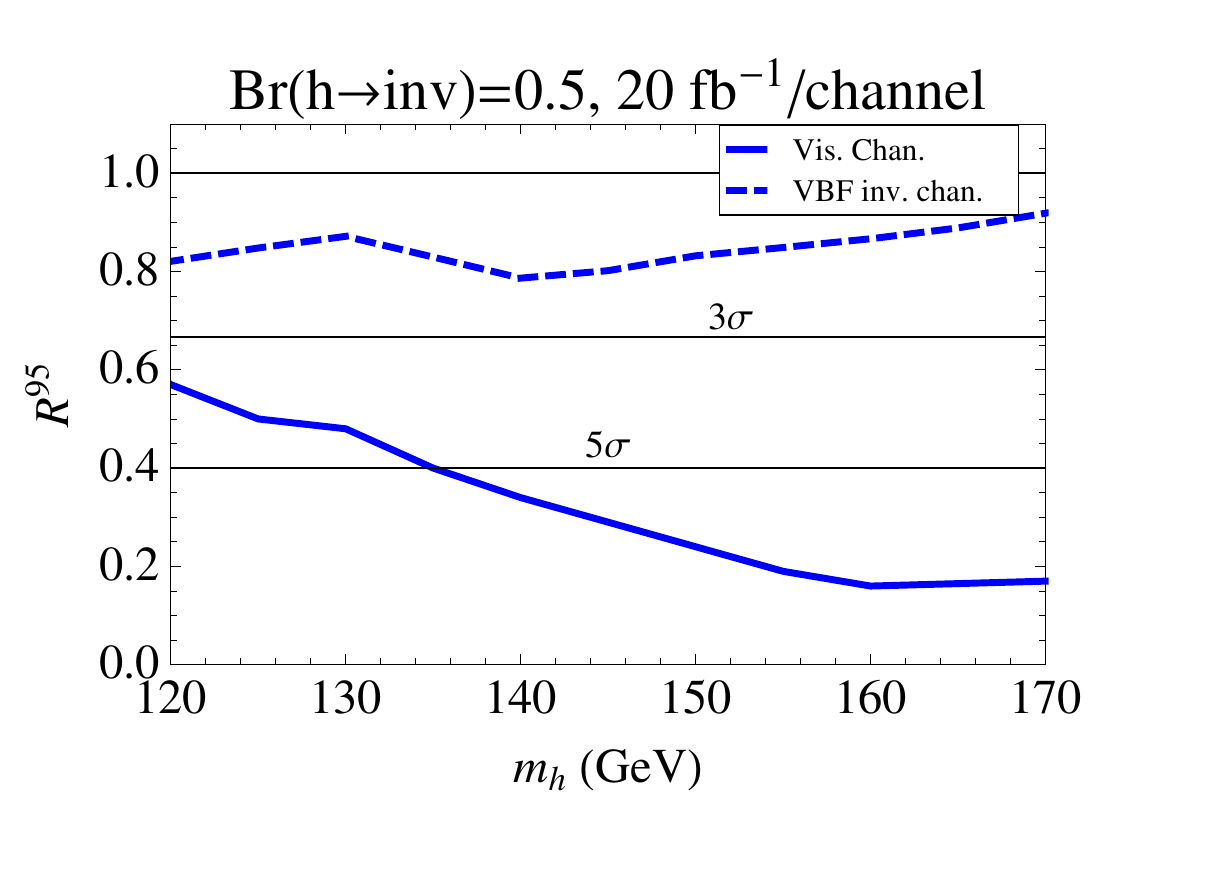} 
\includegraphics[width=0.45\textwidth]{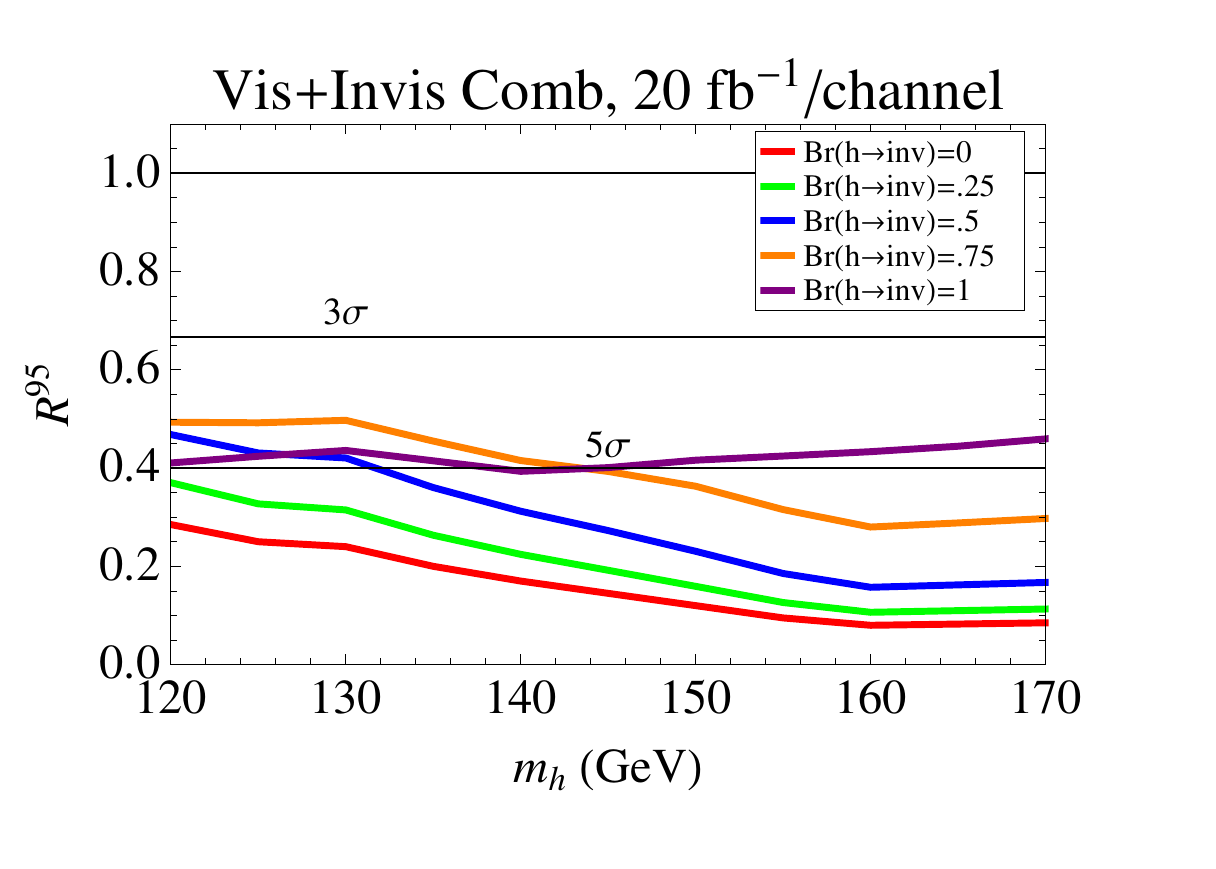} 
\caption{Left: 8 TeV expected CMS visible and WBF invisible 95\% CL reaches
  for the IHH with 50\% branching to invisibles. Right: 8 TeV expected CMS
  visible+WBF invisible combined reach for several $h\rightarrow$ invisible
  branching fractions. Horizontal lines indicate where sensitivity
  reaches 3 and 5$\sigma$; curves below the lines indicate models that
  can either be strongly excluded, or would be expected to give enough
  signal for evidence or observation.}
\label{fig:combsig}
\end{center}
\end{figure*}

The expected limits obtained from a combination of
CMS visible channels and the WBF invisible channel at 20 fb$^{-1}$/channel and 8 TeV are
shown in the right panel of Fig.~\ref{fig:combsig} as a function of
$m_h$ and $BR(h\rightarrow \mbox{inv})$. We use only the WBF channel;
in principle the reach could be extended using monojet and in
particular $hZ$ in combination.  The left panel of Fig.~\ref{fig:combsig}
shows the reach for $BR(h\rightarrow \mbox{inv}) =0.5$ without channel combination,
demonstrating the increase in sensitivity afforded by the combination,
particularly at low Higgs masses.

For low Higgs masses, the invisible reach is increasingly comparable to the visible channels. If the expected reach in invisible channels
with $BR(h\rightarrow \mbox{inv})=1$ is comparable to the expected reach in
visible channels with $BR(h\rightarrow \mbox{vis})=1$, then the combined
expected reach for mixed branching ratios is weaker -- for this reason
$BR(h\rightarrow \mbox{inv})=0.5$ at $m_h=120$ GeV has lower significance
than either $BR(h\rightarrow \mbox{inv})=0$ or $BR(h\rightarrow \mbox{inv})=1$ at
the same mass.

Similarly, if a signal is seen in the SM Higgs search channels, but
the best-fit rate is less than expected from the SM, the invisible
channels will be relevant. If a signal is observed in these channels,
then an SM+invisible combination becomes very interesting; if no
invisible signal is detected then this suggests either altered
production cross-sections or different beyond-the-SM decay modes.

\section{Conclusions and Discussion}
\label{sec:conclusions}

If a Higgs-like particle is observed at the LHC, measuring its
properties in detail is important for understanding the emerging
description of electroweak symmetry breaking.  New physics beyond the
Standard Model can distort the Higgs production cross-sections and
branching fractions.  The branching fractions of a light Higgs
($m_h\lsim 2 m_W$, and in particular $m_h\sim 125$ GeV) are
particularly sensitive to the existence of new light degrees of
freedom, owing to the small SM width of a Higgs below the $W$
threshold.  Searching for beyond-the-SM decays of a Higgs boson is
thus an important cross check of Higgs properties.

We have focused here on light Higgs decay into invisible particles, as (1) these
decays are theoretically well motivated by the existence of dark
matter among other considerations, and (2) the signatures of these decays
are relatively clean, allowing sensitivity to branching fractions
$\mathcal{B}_{inv}=\sigma \times BR(h\to~\mbox{inv})/\sigma_{SM} < 1$ already at the 7 and 8 TeV LHC.
We have studied the 7 and 8 TeV LHC reach for invisibly decaying Higgs
bosons. We have obtained the current direct upper limits on
$\mathcal{B}_{inv}$ from experimental searches in $j+\met$ and
$\ell^+\ell^-+\met$ final states, and performed a full detector-level
study of the weak vector boson fusion channel, which is the most
sensitive.  Our analysis of the WBF channel is the first study at 7 and 8
TeV for the signal $h\to$ invisibles.  
We find, after inclusion of the $gg\to h +$ jets contribution to the
signal and arguing for the use of reweighted $\gamma+$ jets to reduce
the systematic errors in the $Z,W+$jets backgrounds, that invisible
Higgs branching fractions as low as $\mathcal{B}_{inv}\gsim 0.4$ can
be excluded with 20 fb$^{-1}$ of data at 7 and 8 TeV for all Higgs masses
below 170 GeV.

Finally, we have considered the complementarity between searches in
visible and invisible channels. If the branching ratios are highly
mixed, then the hypothesis of invisible Higgs decays (as opposed to
direct suppression of the Higgs production cross section) can be
efficiently tested by a combination of the two types of channels.  The
interesting question of establishing whether Higgs production and
decay are indeed SM-like is thus one that the LHC can begin to
usefully address even in the low-energy run.

\subsection*{Acknowledgments} 
We would like to thank H.~Beauchemin, M.~Buckley, L.~Dixon,
T.~Golling, and S.~Hoeche for useful discussions and comments, and
A.~Hook for providing modified PGS code. SLAC is operated by Stanford
University for the US Department of Energy under contract
DE-AC02-76SF00515. PD was supported by DOE grant DE-FG02-04ER41286. JS
was supported by the DOE grant DE-FG02-92ER40704 and by the LHC Theory
Initiative through the grant NSF-PHY-0969510.  JS and YB thank the
KITP where this work was initiated and which receives support from the
National Science Foundation under Grant No. NSF PHY05-51164.  This
work was also supported by the facilities and staff of the Yale High
Performance Computing Center through the NSF grant CNS 08-21132.

\providecommand{\href}[2]{#2}\begingroup\raggedright\endgroup

 \end{document}